\documentclass[twocolumn,showpacs,superscriptaddress,prl,floatfix]{revtex4}
\usepackage{amsbsy,amssymb,amsmath,bm}
\usepackage{graphicx}

\newcommand{\sign}{\mathop{\mathrm{signum}}}

\newcommand{\f}{fluctuator }
\newcommand{\fs}{fluctuators }

\newcommand{\ba}{\mathbf {a}}

\newcommand{\bB}{\mathbf {B}}
\newcommand{\bF}{\mathbf {F}}

\newcommand{\br}{\mathbf {r}}

\newcommand{\cP}{\mathcal {P}}
\newcommand{\cH}{\mathcal {H}}

\newcommand{\cL}{\mathcal {L}}

\newcommand{\cX}{\mathcal {X}}

\usepackage{color}
\newcommand{\ignore}[1]{}
\newcommand{\bComment}[1]{}
\newcommand{\yComment}[1]{}
\newcommand{\dComment}[1]{}
\newcommand{\ds}[1]{}
\newcommand{\yg}[1]{}

 \renewcommand{\bComment}[1]{\textcolor{blue}{Boris: #1}}
 \renewcommand{\yComment}[1]{\textcolor{green}{Yuri: #1}}
 \renewcommand{\dComment}[1]{\textcolor{blue}{Dan: #1}}
\renewcommand{\ds}[1]{ \vspace*{0.05in} {\color{blue}\hrule
    \noindent\textsf{Daniel: #1}\hrule}\vspace*{0.05in} }
\renewcommand{\yg}[1]{ \vspace*{0.05in} {\textcolor{green}\hrule
\noindent\textsf{Yuri: #1}\hrule}\vspace*{0.05in} }

\newcommand{\fig}[1]{Fig.~\ref{#1}}
\newcommand{\eq}[1]{Eq.~(\ref{#1})}
\newcommand{\eqs}[2]{Eqs.~(\ref{#1}) and~(\ref{#2})}
 \def\be{\begin{equation}}
\def\ee{\end{equation}}

\begin{document}
\title{Non-Gaussian low-frequency noise as a source of qubit
decoherence}
\author{Y. M. Galperin}
\email{iouri.galperine@fys.uio.no} \affiliation{Department of
Physics, University of Oslo, PO Box 1048 Blindern, 0316 Oslo,
Norway} \affiliation{ Argonne National Laboratory, 9700 S. Cass
Av., Argonne, IL 60439, USA} \affiliation{A. F. Ioffe
Physico-Technical Institute of Russian Academy of Sciences, 194021
St. Petersburg, Russia}

\author{B. L. Altshuler}
\affiliation{Physics Department, Princeton University,
        Princeton, NJ 08544, USA}
\affiliation{NEC Research Institute, 4 Independence Way,
        Princeton, NJ 08540, USA}
\author{J. Bergli}
\affiliation{Physics Department, Princeton University,
        Princeton, NJ 08544, USA}
\author{D. V. Shantsev}
\affiliation{Department of Physics, University of Oslo, PO Box
1048 Blindern, 0316 Oslo, Norway} \affiliation{A. F. Ioffe
Physico-Technical Institute of Russian Academy of Sciences, 194021
St. Petersburg, Russia}
\date{ 
\today}

\begin{abstract}
We study decoherence in a qubit with the distance between the two
levels affected  by random flips of bistable  fluctuators.  For the
case of a single fluctuator we evaluate explicitly an exact expression
for the phase-memory  decay in the echo experiment with a resonant ac
excitation.  The echo signal as a function of time shows a sequence of
plateaus. The position and the height of the plateaus can be used to
extract the fluctuator switching rate $\gamma$ and its coupling
strength $v$.  At small times the logarithm  of the echo signal is
$\propto t^3$.  The plateaus disappear when the decoherence is induced
by many fluctuators. In this case the echo signal depends on the
distribution of the fluctuators parameters. According to our analysis,
the results significantly deviate from those obtained in the Gaussian
model as soon as $v \gtrsim  \gamma $.

\end{abstract}


\maketitle

\paragraph{Introduction:}

Quantum dynamics of  two-level systems has recently attracted special
attention in connection with ideas of quantum information processing.
The central problem regarding operation of qubits and logical gates is
maintaining the   phase coherence in the presence of a noisy
environment~\cite{NiChu}.  At low temperatures the noise is dominated
by discrete sources, it is caused by random charge exchange between
localized states and electrodes in the Josephson~\cite{Nak} or
semiconductor double quantum-dot qubits~\cite{Hayashi}. The charge
fluctuations are often modeled by a set of harmonic oscillators with
certain frequency spectrum \cite{Leggett,Weiss}. In these
``spin-boson'' models the qubit decoherence is determined solely by
the \emph{pair correlation function} of random forces, $S_{\cX}(f)$,
that implicitly assumes the noise to be Gaussian \cite{Shnirman}. This
assumption however does not hold
in most practical systems where $S_{\cX}(f) \propto 1/f$ and the
processes have extremely broad distribution of the relaxation
times~\cite{Kogan}.

To understand the role of the non-Gaussian statistics, we
follow~\cite{history} and model the environment  by a set of two-state
systems (fluctuators) that randomly switch between their states.
Their nonequilibrium dynamics can then be taken into account
explicitly as was done in the analysis of  coherent quantum transport
in the presence of $1/f$-noise~\cite{GaChao1}.  Recent application of
a similar approach  to qubits demonstrated new features in the
decoherence that are  not reproduced in the Gaussian
approximation~\cite{Paladino}.  Quantum aspects of non-Markovian
kinetics were addressed in~\cite{Loss}.

In the present paper we extend the work \cite{Paladino} in two
directions. Firstly, we evaluate explicitly the phase-memory decay in
the {\em echo} experiment.  We find a pronounced non-Gaussian behavior
and explain plateaus observed in the time dependence of echo
signal~\cite{Nak}.  Secondly, we consider the case where the
interaction strengths between the qubit and \fs are broadly
distributed.  This distribution strongly modifies the time dependence
and smears away the plateaus.  We suggest a recipe for extracting the
fluctuators' parameters from the measured echo signal.

It is worth noting that  a broad distribution of fluctuators'
switching rates and the coupling strengths makes the problem similar
to the conventional models of the \emph{spectral diffusion}. It was
introduced by  Klauder and Anderson~\cite{KlauderAnderson} for the
problem of spin resonance.  Black and Halperin~\cite{BlackHalperin}
generalized it ~\cite{KlauderAnderson} to phonon echo and saturation
of sound attenuation by two-level systems in
glasses~\cite{AHVP}. These ideas further developed
in~\cite{HuWalker,Laikhtman} were applied to the single molecular
spectroscopy in disordered media~\cite{SMS}.

\paragraph{Model:}
\begin{figure}[b]
\centerline{ \includegraphics[width=6cm]{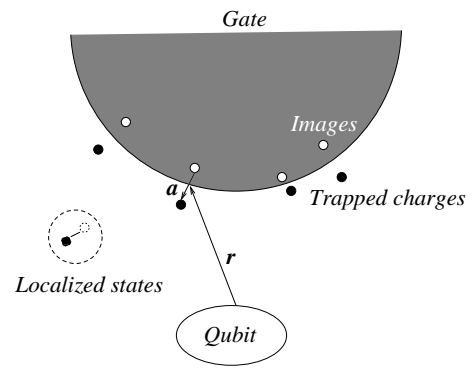} }
\caption{Schematic distribution of charged traps located  near the gate
surface and producing oppositely charged images. \label{fig1}}
\end{figure}
We assume that the qubit is a two-level system (TLS) surrounded by \fs
--  systems with two locally stable states.  Possible candidates for
such fluctuators  in solid state devices are charge traps, see
\fig{fig1}, or structural dynamic defects. An occupied trap together
with its charge image produces a dipole electric field fluctuating in
time due to hops between the trap and the gate and acting upon the
qubit. If the  defects are not charged, they can behave as elastic
dipoles producing time-dependent strains and interacting with the
qubit via deformational potential.

A qubit coupled to the environment will be modeled by the Hamiltonian
$\tilde{\cH}=\tilde{\cH}_q +\tilde{\cH}_\text{man}+\tilde{\cH}_{qF}
+\tilde{\cH}_{F}$, where $\tilde{\cH}_q$ and $\tilde{\cH}_F$ describe
the qubit and the fluctuators separately. A completely isolated qubit
has two states and is characterized by the  energies of these states
and their tunneling coupling amplitude.  $\tilde{\cH}_q$ is the
Hamiltonian of the qubit pseudospin $\bm{\sigma}$ in a static
``magnetic field", $\bB =\left\{B_x,B_z\right\}$. Here $B_z$
characterizes the splitting of the energies of the two states, and
$B_x$ describes their tunneling coupling. This Hamiltonian can be
diagonalized by rotation in the pseudospin space  with new $z$-axis
parallel to $\bB$. The rotated Hamiltonian, $\cH_q$, is then $\cH_q =-
(B/2)\sigma_z$. $\tilde{\cH}_F$ can be diagonalized by a similar
rotation in the fluctuator's pseudospin space and then  split into
three parts, $\cH_F= \cH_{F}^{(0)}
+\cH_{F\!\!-\text{env}}+\cH_{\text{env}}$.  The first part is just a
Hamiltonian of an isolated two-level tunneling system, $
\cH_{F}^{(0)}=\sum_i (E_i/2) \tau_z^{i}$, where the Pauli matrices
$\bm{\tau}^{(i)}$ correspond to $i$-th fluctuator. The spacing between
the two levels, $E_i$, is formed by the diagonal splitting, $\Delta_i$,
and the tunneling overlap integral, $\Lambda_i$, as
\begin{equation}
  \label{eq:thetai}
  E_i=\sqrt{\Delta_i^2 + \Lambda_i^2}\equiv \Lambda_i/\sin \theta_i \,
  .
\end{equation}
The flip-flops of the \fs are due to the coupling with a thermal bath
which we model by a system of  equilibrium bosons. This applies to
phonons, as well as  to electron-hole pairs in conducting part of the
device~\cite{Black}.

The interaction, ${\cH}_{qF}$, between the qubit and the fluctuators
is specified as (cf. with Ref.~\onlinecite{BlackHalperin})
\begin{equation}
  \label{eq:004b}
 \cH_{qF}=\sum_i v_i \,
 \sigma_z\tau_z^i\, , \quad  v_i=u(r_i)
 \cos \theta_i \, .
\end{equation}
Here we assumed for simplicity that the coupling strength $v_i$ is
determined only by $\theta_i$ defined by Eq.~(\ref{eq:thetai}) and the
distance  $r_i$ between the qubit  and the $i$th fluctuator.

The interaction between the \fs and the environment manifests itself
through time-dependent random fields applied to the qubit. Frequencies
of these fields being much smaller than the temperature $T$ and qubit
splitting $B$,the fields can be treated \emph{classically}:
$\hat{\tau}^{(i)}_z \to \xi_i(t)$. Accordingly, $\cH_{qF}$ is the
Hamiltonian of the qubit pseudospin in a  random, time-dependent
magnetic field $\cX(t)$ formed by independent contributions of
surrounding fluctuators:
\begin{equation}  \label{tp4}
\cH_{qF}=\cX(t)\, \sigma_z\,  , \quad
  \cX(t)=\sum_i v_i \xi_i(t) \, .
\end{equation}
The random functions $\xi_i(t)$ characterize the fluctuators' state:
$\xi_i(t)$ instantly switches between $\pm 1/2$ at random times
(random Poissonian process).  The switching rates, $\gamma_i$, can be
calculated in the second order of the perturbation theory for the
fluctuator-phonon/electron interaction~\cite{Jackle,Black},
\begin{equation}
\label{eq:007}
 \gamma_i = \gamma_0(T) \sin^2 \theta_i\,.
 \end{equation}
$\gamma_0$ is thus the \emph{maximal} fluctuator switching  rate at a
given temperature, $T$. For simplicity we assume here that the \f
spends on average an equal time in each state.  Although justified
only for $E_i \ll T$, this assumption produces correct temperature
dependences~\cite{Laikhtman}.

The qubit is manipulated by applied ac ``magnetic field'', $\bF
(t)\parallel \mathbf{x}$, with frequency close to  $B$, so that
$\cH_\text{man}=(1/2)F(t)\sigma_x$.  The echo-like manipulation allows
substantial suppression of the decoherence comparing to the ``free
induction'' signal decay~\cite{Nak}.  In this case a resonant ac field
is first applied as a pulse rotating the qubit's pseudospin by $\pi/2$
($\pi/2$-pulse).  After a delay $\tau$,  a $\pi$-pulse inverting the
qubit's pseudospin is applied, and the echo $\pi/2$-pulse appears
after another delay $\tau$. After the $\pi$ pulse the phase evolves in
the reverse direction, thus only actual switchings during the time
$2\tau$ contribute to the decoherence, while  static \fs do not affect
the echo signal.

The external pulses are usually short enough for both relaxation
and spectral diffusion during each of the pulses to be neglected.
The echo decay is known to be proportional to the ``phase-memory
functional"~\cite{Mims}
\begin{equation}
\psi=\left \langle e^{i\varphi_\tau} \right \rangle_{\!\xi_i}\, ,
\quad \varphi_\tau \equiv \int_0^{2\tau} \beta(t',\tau) \cX(t')\,
dt' \label{eq:016}
\end{equation}
where $\beta(t',\tau) \equiv \sign (\tau -t')$.
The average is calculated over the realizations of the random
processes $\xi_i(t)$ and random initial states of \fs \!.  This
averaging reflects the experimental procedure where the observable
signal is an accumulated result of numerous repetitions of the
same sequence of inputs.  Equation~(\ref{eq:016}) is obtained by
analysis of the qubit's density matrix under the perturbation
$\cH_{\text{man}}$.

\begin{figure}[h]
 \centerline{\includegraphics[width=7cm]{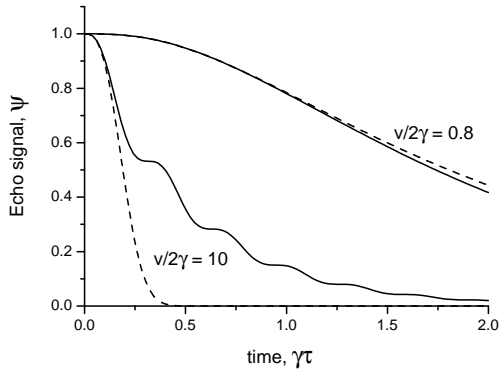} } \caption{Echo
signal for different values of the ratio $v/2\gamma$ (shown near the
curves), \eq{eq:04}. Dashed lines - calculations along the Gaussian
assumption, \eq{Gauss}.
\label{fig:04}}
\end{figure}

\paragraph {Single fluctuator: }
Let us start analyzing \eq{eq:016} with the case of a qubit
interacting with a single fluctuator. In the spirit of
Ref.~\onlinecite{Laikhtman} we have obtained the exact solution
for the echo signal~\cite{our},
\begin{equation}
 \psi=\frac{e^{-2\gamma\tau}}{2\mu^2}\left [ (1+\mu)
 e^{2\mu\gamma\tau}+(1-\mu) e^{-2\mu\gamma\tau} -
 \frac{v^2}{2\gamma^2}  \right]  \label{eq:04}
\end{equation}
where  $\mu= \sqrt{1-(v/2\gamma)^2}$. This result
is essentially non-Gaussian. Indeed, assuming that the phase,
$\varphi_\tau $,
obeys the Gaussian statistics one would get instead of \eq{eq:04}
$\psi_G=e^{-\langle \varphi^2_\tau \rangle/2}$ with
\begin{equation}
\langle \varphi_\tau^2\rangle/2
= (v/4\gamma)^2\left(4\gamma\tau -3 +4e^{-2\gamma \tau}
- e^{-4\gamma \tau} \right)\, . \label{Gauss}
\end{equation}
Comparing the two expressions we notice that the Gaussian result
(\ref{Gauss}) is the weak coupling limit, $v \ll 2\gamma$, of the
exact solution (\ref{eq:04}).  However, in the strong coupling
case, when $v \gtrsim 2 \gamma$,
the exact solution strongly differs from Eq.~(\ref{Gauss}). In
Fig.~\ref{fig:04} both functions are plotted for
different values of the ratio $v/2\gamma$.
One can see that at $v \ge 2\gamma$ the Gaussian assumption strongly
underestimates the phase-memory functional for $\tau >
2\pi/v$. Similar conclusion for the free induction signal has been
recently obtained  in Ref.~\onlinecite{Paladino}. The reason for the
failure of the  Gaussian approximation in the strong coupling case is
similar to the  well known motional narrowing of spectral
lines~\cite{KlauderAnderson}. Indeed, if $v \ge 2\gamma$, then each
fluctuator \textit{splits} the qubit's levels rather than broadens
them. The qubit just experiences rare hops between these states; the
splitting is of the order of $v$ and the typical hopping rate is
$\gamma$.  On the contrast, at $v\ll \gamma$ the splitting between the
levels is smeared, the typical decay rate of the echo signal being
$\sim v^2/\gamma$. This limit is reproduced  within the Gaussian
assumption. At $\gamma \tau \ll 1$, the phase-memory functional
behaves as
\be
\psi \approx 1 -  \gamma v^2 \tau^3/3\, , \label{t3} \ee
regardless of the value of the ratio $v/\gamma$.  This result naturally holds
also in the Gaussian approximation. At
$\gamma \tau \gg 1$ we find
$$-\ln \psi \approx \left\{ \begin{array}{ll} 2 \gamma \tau,  & v >
 2\gamma \, ;\\ v^2 \tau /4\gamma, &  v\ll 2\gamma \, .\end{array}
 \right.$$  At $v > 2\gamma$ there appear steps in the time dependence
 of the echo signal shown in Fig.~\ref{fig:04}. It looks like these
 have been experimentally observed in Ref.~\onlinecite{Nak} (see
 Fig. 3 there). At $v \gg \gamma, \sqrt{\gamma/\tau}$
 Eq.~(\ref{eq:04}) acquires a simple form
 \be
 \psi=e^{-2\gamma\tau}\left[1+(2\gamma/v)\, \sin v \tau \right]\, .
\label{plat}
\ee
According to Eq.~(\ref{plat}), the plateau-like features
($d\psi/d\tau \approx 0$) occur at $v \tau \approx 2 k
\pi$ and their heights $\psi\approx e^{-4\pi k\gamma/v}$ exponentially
decay  with the number $k$. These plateaus for $\gamma
\ll v$  can be understood as follows.  The probability for the
fluctuator to flip during the time $2\pi/v$ is very small, hence it
either does not flip at all or flips only once.  If it flips during
the first half of the beating period, $t<\pi/v$, the phase of the
functional (\ref{eq:016}) at $2\tau=2\pi/v$  evolves from $0$ to
$\pi$. If it flips   during the second half of the period,
$\pi/v<t<2\pi/v$, the phase evolves from $\pi$ to $2\pi$. After
averaging the two contributions will cancel each other, which implies
that at $\tau$ close to $2\pi/v$ the signal $\psi$ is almost
insensitive to small variations of $\tau$.

Measuring experimentally  the position and the height of the first
plateau, one  can determine both the fluctuator coupling strength $v$,
and its switching rate $\gamma$.  For example, the echo signal
measured for a Josephson junction qubit in Ref.~\onlinecite{Nak} shows
a plateau-like   feature at $\tau=3.5$~ns at the height
$\psi=0.3$.
$\gamma \approx 27$~MHz.  If the fluctuator is  a charge trap near
the gates  producing a dipole electric field, see Fig.~\ref{fig1},
its coupling strength is $v=e^2(\ba\cdot \br)/r^3$.  Estimating the
actual gate-qubit distance $r\approx 0.5\ \mu$m, we obtain
a reasonable estimate for the tunnelling
distance between the charge trap and the gate, $a\sim 20$ \AA.

\paragraph{Many fluctuators:}
What if the qubit is coupled to many fluctuators, which are not
correlated: $\langle \xi_i(t) \xi_j(t') \rangle \propto \delta_{ij}$?
The phase memory functional is then a product of the partial
functionals due to individual fluctuators,  $\Psi = \prod_i
\psi^{(i)}$. Following  the Holtsmark
approach~\cite{KlauderAnderson,Laikhtman,our} we approximate $\Psi$ as
$\Psi \approx \exp[-\sum_i (1- \psi^{(i)})]$.  Our approach~\cite{our}
provides accurate description of the decoherence by fluctuators with
particular locations as long as the number of active \fs is large.

We will show now that summation over many fluctuators with a broad
distribution of $v_i$ and $\gamma_i$ may significantly change the
time-dependence of the echo signal.  Since the tunneling splitting,
$\Lambda$, depends exponentially on the distance in real space between
the positions of the two-state fluctuator, it is reasonable to assume
that the distribution function $\cP(\Delta,\Lambda) =\eta/{\Lambda}$.
In terms of $E$ and $\theta$ it implies $\cP(E, \theta) = \eta/\sin
\theta$. Here $\eta$ is proportional to the number of fluctuators with
$E  \lesssim T$, which are not frozen at a given temperature. Hence,
$\eta \propto T$.  Note also that the contribution of one \f to the
noise spectral density $S_\cX(\omega)$ is proportional to
$\gamma/(\omega^2+\gamma^2)$, hence the noise spectrum in the
frequency window   $\gamma_{\min} \ll \omega \ll \gamma_0$ is of the
$1/\omega$ type.

Due to existence of a finite maximal rate, $\gamma_0$, summation over
fluctuators with different $\gamma_i$ does not affect much the
decoherence at small times. Similarly to \eq{t3} we find that $-\ln
\Psi \propto \tau^3$ for $\gamma_0 \tau\ll 1$. This asymptotic
behavior can however be modified due to a distribution of the coupling
strengths $v_i$ entering \eq{eq:004b}. Let us assume that $u(r)\propto
r^{-2}$ which is the case if the fluctuators act as electric
dipoles. Uniform distribution of the \fs  over a two-dimensional gate
surface corresponds to the distribution of the coupling constants
$\cP(u) \propto u^{-2}$, which coincides with the distribution of the
coupling constants in glasses, where two-level systems interact via
dipole-dipole interaction~\cite{BlackHalperin}.  The phase-memory
functional can  be evaluated by  integrating the echo signal $\psi$,
(\ref{eq:04}), where $\gamma$ and $v$ are related to $u$ and $\theta$
by \eqs{eq:004b}{eq:007},
 \begin{equation}
   \Psi=\exp\left[ - u^* \int_{0}^{\infty} \frac{du}{u^2}\int_0^{\pi/2}
   \frac{\left[1-\psi (u,\theta)\right] \ d \theta}{\sin   \theta}
      \right]\, .
   \label{int}
 \end{equation}
Here $u^*$ is the value of $u(r)$ taken at the average distance
between the fluctuators having  energy splitting $\lesssim T$. The
integration over $u$ can be extended down to zero  since $(1-\psi)
\propto u^2$ at $u\to 0$.
The combination $\cL \equiv -(\gamma_0/u^*)\ln \Psi$ turns out to be
an universal function of the product $\gamma_0\tau$.
At small time
 $\cL \approx (\pi/4)(\gamma_0 \tau)^2$. A stronger time
dependence compared to the single-fluctuator case, \eq{t3}, is due
to strongly-coupled fluctuators with large $v$. In the general
case, when $u(r)\propto r^{-b}$, and the fluctuators are uniformly
distributed over an area of dimension $d$, one finds $ \cL \propto
\tau^{1+d/b}$ for $ \gamma_0 \tau\ll 1$.  Comparing the short-time
experimental dependence with this prediction one can estimate the
ratio $d/b$, i. e., extract an information on the spatial distribution of
fluctuators  and mechanism of their interaction with the qubit.
One should also keep in mind that energy relaxation described by
the characteristic time $T_1$ provides an additional echo decay as
$\propto e^{-2\tau/T_1}$, see~\cite{ast}. We neglect this
contribution to the echo decay.

In reality, the distribution over $u$ should be cut off at some
$u_{\max}$ corresponding to the minimal possible distance between
the fluctuator and the gate. Our selection of the relevant \fs (see
above) is valid provided that $u_{\max}/\gamma_0 \gg 1$.
At $\gamma_0 \tau \gg 1$ one obtains $\cL \propto
\tau$~\cite{our}.

The presented results can be used to probe the sources of qubit
decoherence by measuring time dependence of the echo signal. The
plateaus on this dependence signals that a major
contribution comes from a \textit{single} fluctuator, which parameters
can be extracted from the position and height of the plateaus using
\eq{plat}. A smooth time dependence of $\ln \Psi$ suggests a combined
effect of many fluctuators. These conclusions are qualitatively
correct also for the ``free-induction'' signal.

Many fluctuators with a broad distribution of relaxation rates produce
noise with $1/f$-type spectrum.  It is worth mentioning, see
Ref.~\onlinecite{our} for details, that generally phase memory decay
and $1/f$ noise are dominated by \textit{different} groups of
fluctuators. Consequently, in general case the fluctuator-induced
decoherence \textit{cannot} be expressed through the noise
spectrum. This is in contrast to the statements that often appear in the
literature. Non-Gaussian effects due to a single or many \fs may be
responsible for the experimentally observed
decoherence~\cite{Nak,lastprep}.

\acknowledgments {This work was partly supported by
FUNMAT\verb+@+UiO, Norwegian Research Council, and by the U. S.
Department of  Energy Office of Science through contract No.
W-31-109-ENG-38. We are thankful  to V. Kozub, I.~Lerner and  V. Kravtsov
for discussions of theoretical issues and Y.~Nakamura, J. S. Tsai,
Yu. A. Pashkin, and O. Astafiev for discussions of experimental
aspects.}

\end{document}